\documentclass[reprint,eqsecnum,floats,aps,amsmath,amssymb,nofootinbib,prd,onecolumn,showpacs]{revtex4-2}

\usepackage{graphicx}
\usepackage{bm}
\usepackage{amsmath}	
\usepackage{amsfonts}
\usepackage{mathtools}
\usepackage{amssymb}							
\usepackage{braket}
\usepackage{amsthm}			
\usepackage{hyperref}

\usepackage[T1]{fontenc} 

\usepackage{color}

\begin{document}

\title{Starobinsky potential and power suppression in hybrid Loop Quantum Cosmology}

\author{Marceau Henry}
\email{marceauhenry5@gmail.com}
\affiliation{Instituto de Estructura de la Materia, IEM-CSIC, Serrano 121, 28006 Madrid, Spain}

\author{Guillermo A. Mena Marug\'an}
\email{mena@iem.cfmac.csic.es}
\affiliation{Instituto de Estructura de la Materia, IEM-CSIC, Serrano 121, 28006 Madrid, Spain}

\author{Antonio Vicente-Becerril}
\email{antonio.vicenteb@estudiante.uam.es}
\affiliation{Instituto de Estructura de la Materia, IEM-CSIC, Serrano 121, 28006 Madrid, Spain}

\begin{abstract}
We study the effect on the primordial tensor power spectrum of varying the number of e-folds during slow-roll inflation in Loop Quantum Cosmology with a Starobinsky potential. Using the hybrid quantization approach, we derive the effective mass governing tensor mode evolution. The choice of vacuum state is crucial, especially since the preinflationary phase predicted by Loop Quantum Cosmology invalidates the choice of the Bunch-Davies state as the preferred vacuum. We adopt a choice which is optimally adapted to the dynamics, so that it provides a non-oscillating (NO) spectrum free of spurious contributions, and permits an asymptotic Hamiltonian diagonalization (AHD) of the perturbations. For this so-called NO-AHD vacuum, we compute the power spectrum using both analytic approximations and numerical integration. Our results confirm the accuracy of our approximations in a wide range of situations, including short- and long-lived inflationary scenarios. The primordial power spectrum exhibits a characteristic cutoff on a wavenumber scale determined primarily by the background dynamics around the bounce that replaces the big bang in Loop Quantum Cosmology.
\end{abstract}

\maketitle

\section{Introduction}

In the last decades, the observation of the Cosmological Microwave Background (CMB) and its confrontation with theoretical predictions has become increasingly important thanks to the improved precision of the measurements completed, for instance, by collaborations such as \textit{WMAP} and \textit{Planck} \cite{WMAP,Planck_anomalies,Planck_2018}. The observational data have revealed certain tensions or indicated the possibility of certain anomalies with respect to the standard inflationary models (see e.g. Refs. \cite{Planck_anomalies,ACT}). In particular, it has been suggested that there is a statistically unexpected low level of correlation at large angles and a lack of power in the low multipole moments \cite{tango}. As a possible mechanism to (at least) alleviate these problems, some authors have proposed models where the primordial power spectrum (PPS) exhibits a cutoff at large angular scales \cite{cutoff1,cutoff2,cutoff3,cutoff4}. 

An appealing explanation for the appearance of an (effective) cutoff may be found in quantum gravity phenomena \cite{Maciej}. In this context, Loop Quantum Gravity (LQG) has emerged as a popular candidate for a quantum version of general relativity. LQG is a nonperturbative canonical quantization of the Einsteinian theory formulated in terms of su(2)-connections and densitized triads (the so-called Ashtekar-Barbero variables) \cite{LQG, Thie}. When the techniques of LQG are applied to symmetry reduced spacetimes, as in cosmological models relevant to the CMB, a new discipline known as Loop Quantum Cosmology (LQC) arises \cite{LQC}. One of the most remarkable achievements of LQC is the resolution of the standard cosmic singularity,  which is replaced by a quantum bounce, commonly referred to as the \textit{big bounce} \cite{APS,APS1}. This provides a mechanism for a quantum transition between a contracting universe to an expanding one.

The application of LQC to the study of primordial perturbations in cosmology and the extraction of potentially observable predictions have received considerable attention in recent years \cite{effective2,revAWW,revAB,hybrid_rev}. To carry out this study, a necessary step is to evolve the cosmological perturbations through the cosmic preinflationary and inflationary epochs. In the LQC literature, cosmological perturbations have been analyzed employing different quantization approaches, such as the dressed metric approach \cite{AAN,AM,dressed1,dressed2,dressed4}, the hybrid quantization approach \cite{hybridCMB,hybr_inf1,hybrid_rev,hybr_inf2,hybr_ten}, or the anomaly-free approach \cite{anomalyfree1,anomalyfree2,effective4,effective5,anomalyfree3,anomalyfree4}. The first two of these approaches respect the hyperbolicity of the equations of the perturbations in the ultraviolet sector and do not distort the relativistic dispersion relations. In both of these cases, it has been possible to show that LQC can explain a power suppression in the PPS \cite{tango,NM}. In the present work, we are going to focus exclusively on the hybrid approach, in which some preliminary works have detected higher rates of suppression around the effective cutoff scale \cite{MVY,MVY2}. However, our analysis can also be extended to the dressed metric approach almost straightforwardly by appropriately modifying the effective mass term, as shown in Refs. \cite{Simon,MVY}. Similarly, our framework can also be applied to scalar perturbations by using the corresponding effective mass for scalar modes during the inflationary period \cite{MVY2}. In this sense, we can safely assume that our conclusions can be generalized to these other situations.

To determine the solution to the propagation equations of the tensor modes, one needs to know the initial conditions on these perturbations. In cosmology, the fixation of the initial state of the perturbations is akin to fix the choice of a vacuum. In standard inflationary cosmologies, such as approximately de Sitter spacetimes, the Bunch Davies state \cite{Bunch,Mukhanov1} is the unique state which is invariant under the symmetries of the de Sitter background and has a good local behavior, in the sense that it is a Hadamard state \cite{WaldH}. Nevertheless, in models with a preinflationary (non de Sitter) period which significantly affects the modes that are physically observable, the Bunch-Davies state ceases to be a natural choice of vacuum state, because it is not optimally adapted to the relevant background evolution. We recall that, in generic curved non-stationary spacetimes, like those that are found in the aforementioned situations, there is no preferred choice of vacuum state based on symmetry invariance. Furthermore, adiabatic procedures are questionable in scenarios with not ignorable evolution rates. So, the choice of a vacuum supported by physically reasonable criteria becomes a crucial point in the study of the perturbations and the extraction of predictions about them \cite{NM,NBM}.

With the aim of selecting a preferred quantum state as a vacuum for the primordial perturbations, several proposals have been developed in the context of LQC  \cite{Agullo1,Lueders,Handley,AG1,AG2,deBlas,NMT}. In this way, it has become evident that this issue has a significant impact on the PPS in LQC \cite{NBM}. In particular, for generic vacuum choices, which are not optimally adapted to the background dynamics, the spectrum exhibits an oscillatory behavior, especially for modes with small wavenumbers, which are those most affected by the typical scales of the preinflationary evolution. In this work we adopt a proposal that, however, leads to a non-oscillating (NO) PPS, free of the power pumped out in average by such spurious fluctuations. Furthermore, this proposal selects a quantum state with modes that ensure an asymptotic Hamiltonian diagonalization (AHD) of the perturbations in the ultraviolet regime, eliminating mode interactions in this sector. The vacuum state determined with this procedure is therefore often called the NO-AHD vacuum \cite{NMT}.

The inflationary period plays a crucial role in the evolution of the cosmological perturbations until they cross the horizon and freeze. This inflation, driven by the potential of the inflaton, is considered to last, at least, approximately 60 e-folds since the moment in which the largest angular scales directly observable nowadays crossed the horizon \cite{WMAP,Planck_anomalies,efolds}. In short-lived inflationary scenarios, we will take for definiteness and concreteness around 70 e-folds of accumulated expansion at the end of inflation. In this work, by short-lived inflation we mean inflationary cosmologies where one has just enough total e-folds until the end of inflation to achieve consistency with observations. Sometimes this situation is called just-enough inflation in the literature \cite{justenough1,justenough2}.\footnote{For comparison, in our discussion we will also consider a long-lived inflationary case where inflation produces more than a million e-folds.} To reach this number of e-folds, a big number of different potentials have been investigated \cite{Encyclopedia}. Between them, the Starobinsky potential \cite{Starobinsky} is one of most studied ones, because it can be motivated from higher-curvature terms and can be argued to be favored\footnote{Recently, some possible tensions have been claimed to exist, in the light of recent measurements at millimeter wavelengths \cite{ACT}, although possible explanations have been suggested based on a slightly large number of e-folds during reheating \cite{Zharov}.} by observations \cite{Planck_2018,Planck_2015}. In the specific case of LQC scenarios, the evolution of the cosmological background in the presence of this potential has been studied in several works \cite{BG1,BG2,Staro_LQC_1,Staro_LQC_2,Staro_LQC_3,Staro_LQC_4,Staro_LQC_5},
In Ref. \cite{Staro_LQC_1} initial conditions on the inflaton leading to the desired number of e-folds were found, and a preliminary analysis of the perturbations (with a different choice of vacuum state than the one selected here) supported the emergence of an effective cutoff in the PPS.

Our main goal in this work is to explore the accuracy, validity, and robustness of various analytic approximations used to compute the (tensor) PPS of the NO-AHD vacuum state in LQC. We show that, when applied to the Starobinsky potential within the hybrid LQC framework, these analytic approximations yield a PPS that closely matches the one obtained by a full numerical integration. This result supports that the approximations are valid well beyond the case of a quadratic potential, the only inflaton potential for which they have been numerically checked until now in the literature \cite{MVY,MVY2}. To further elucidate the validity of our approximations, we consider different cosmological scenarios by varying the initial conditions, resulting in both short- and long-lived inflationary phases. In both cases, the analytic expressions for the PPS display an excellent agreement with the numerically computed spectra, confirming the consistency and reliability of the approximations.

The structure of the paper is as follows. In Sec. II, we briefly review the dynamical equations of the background and present the main results of the hybrid quantization for the tensor perturbations. We also define the NO-AHD vacuum, which is determined by appropriate initial conditions for the perturbations. Since the mode equations cannot be solved exactly, in Sec. III we introduce suitable approximations to the effective mass, allowing for analytic solutions of the mode equations. Then, in Sec. IV,  we compute the PPS, both analytically and numerically, comparing also the spectra obtained with different initial conditions for the inflaton at the bounce. We analyze the similarities and differences between all these spectra and demonstrate the wide range of applicability of the analytic approximations, which yield a consistent PPS for both short- and long-lived inflation. Finally, Sec. V contains the conclusions. In the following, we use Planck units, setting $G$, $c$, and $\hbar$ equal to one. 

\section{The framework}

In this section, we summarize the most relevant results about the dynamics of the effective background in LQC. We also present the effective mass in the hybrid quantization approach, and determine a vacuum state optimally adapted to the LQC background dynamics (at least around the bounce) by using the NO-AHD vacuum proposal.

\subsection{Background}

We consider a homogeneous and isotropic background described by a spatially flat Friedmann-Lemaître-Robertson-Walker (FLRW) metric, characterized by a scale factor $a(t)$, in the presence of a homogeneous scalar inflaton field $\phi(t)$ subject to a Starobinsky potential $W(\phi)$. This potential is defined as
\begin{equation}
    W(\phi) = \frac{3m^2}{32\pi } \left(1 - e^{-\sqrt{\frac{16\pi} {3}} \phi}\right)^2.
\end{equation}

According to recent observations from the \textit{Planck} mission, we fix the inflaton mass to $ m = 2.51 \times 10^{-6} $ in Planck units \cite{Planck_2015, BG1}. The energy density and pressure of the scalar field are given by  $\rho= \frac{1}{2} \dot{\phi}^2 + W(\phi)$ and $P = \rho - 2W(\phi)$, respectively, where the dot denotes the derivative with respect to proper time. The background dynamics in effective LQC \cite{LQC} is governed by the following modifed Friedmann equation (derived in fact from an effective Hamiltonian) \cite{AAN,hybridCMB}:
\begin{eqnarray}\label{eq_Friedman_LQC}
    \left( \frac{a'}{a}\right)^{2} = \frac{8\pi}{3}a^{2}\rho \left(1 - \frac{\rho}{\rho_{c}} \right).
\end{eqnarray}
The prime denotes the derivative with respect to conformal time and $\rho_c = 3/(8\pi \gamma^2 \Delta)$ is called the critical density. It is the maximum allowed for the energy density in LQC, reached at the big bounce \cite{LQC}. Here, $\gamma$ is the Immizi parameter, fixed to $\gamma=0.2375$ by black hole entropy arguments \cite{immirzi, LQG}, and $\Delta = 4\sqrt{3}\pi\gamma$ is the area gap, derived from the LQG area spectrum \cite{LQG, Thie}. 

The modified Friedmann equation is supplemented with the local conservation law for the inflaton energy density:
\begin{eqnarray}\label{law}
    \ddot{\phi} + 3H\dot{\phi }+W,_\phi(\phi)=0,
\end{eqnarray}
where $H=\dot{a} /a$ is the Hubble parameter and the comma followed by $\phi$ denotes the derivative with respect to the inflaton. In this work, we numerically integrate Eqs. \eqref{eq_Friedman_LQC} and \eqref{law} to obtain the full background evolution, starting from the bounce, as a naturally distinguished time, until the end of inflation. For concreteness, we take the scale factor equal to one at the bounce, $a_{\text{B}} = 1$. The subindex $\text{B}$ stands for evaluation at the bounce. In the following, we choose the proper time at that moment equal to zero for simplicity.   

The phenomenologically interesting solutions in effective LQC, with short-lived inflation, are known to have a negligible contribution of the inflaton potential to the energy density (at least) in the region around the bounce \cite{NM,NBM}.
\footnote{Generally, kinetic dominated bounces lead to short-lived inflation even beyond LQC (see e.g. the discussion in Ref. \cite{barretin}). On the other hand, this kind of bounces have been argued to be favored in standard LQC with natural probabilistic measures for pre-bounce initial conditions on the background \cite{barretin,barretin2}. Probability distribution functions for the number of e-folds of inflation as a function of the kinetic energy density of the inflaton at the bounce have also been studied in Ref. \cite{barretin3}.} Thus, from now on we concentrate our attention on solutions for which the condition \( \frac{1}{2} \dot{\phi}_\text{B}^2 \gg W(\phi_\text{B}) \) is satisfied.  

The derivative of the inflaton at the bounce can be expressed as $ \dot{\phi}_\text{B} = \pm 2 \sqrt{\rho_c - W(\phi_\text{B})}$ in terms of the value of the inflaton. As mentioned in the Introduction, we choose different initial values for the inflaton to obtain either short- or long-lived inflation. To achieve short-lived inflation, we take \( \phi_\text{B} = -1.45 \) with \( \dot{\phi}_\text{B} > 0 \) and \( \phi_\text{B} = 3.63 \) with \( \dot{\phi}_\text{B} < 0 \), both leading to around 70 e-folds at the end of inflation \cite{Staro_LQC_1}. For long inflation, we choose \( \phi_\text{B} = 0.97 \) with \( \dot{\phi}_\text{B} > 0 \), which yields more than one million e-folds. In the three studied situations, the background evolution can be divided into three distinct periods: a bounce period, where quantum geometry effects are significant, a kinetically dominated period, and a slow-roll inflationary phase. This description is consistent with results previously reported in the literature \cite{Staro_LQC_5}. \\

We conclude our numerical integration when the slow-roll inflation ends. This is determined by an internal condition in the integrator, which monitors the slow-roll parameter $\epsilon_H = - \dot{H}/H^2 $. The integrator stops once $\epsilon_H\geq1$,
signaling the end of inflation. This condition is implemented inside the integrator routine and is evaluated from the onset of inflation until the criterion is satisfied. We adopt the explicit Runge–Kutta method of order 8, specifically the \texttt{DOP853} integrator, setting the relative tolerance to $10^{-14}$ and the absolute tolerance to $10^{-12}$.

\subsection{Background-dependent mass in the hybrid approach}

As previously mentioned, in this work we consider the hybrid approach for the quantization of the tensor modes. This approach treats the background and the perturbations as two sectors of a single constrained canonical system. The resulting Hamiltonian constraint is then quantized by applying loop techniques to the background, while a Fock quantization (which is uniquely determined up to unitary transformations) is adopted for the perturbations \cite{hybrid_rev}. For appropriate quantum states (with factorized dependence on the background geometry and the gauge-invariant perturbations), the evolution of the perturbations can be described by effective equations, that resemble those of general relativity except for the replacement of the relativistic background-dependent mass (-$a^{\prime\prime}/a$) by a corrected mass. In the case of tensor perturbations, this mass with quantum corrections, $s^{\text{(t)}}$, takes the form given in Ref. \cite{hybrid_rev}:
\begin{eqnarray} \label{mass_LQC}
    s^{\text{(t)}} = - \frac{4\pi}{3}a^{2} \left(\rho - 3 P \right).
\end{eqnarray}

The tensor modes $\mu_k$ satisfy then the equation
\begin{eqnarray} \label{MS_eq}
    \mu_k'' + \left(k^2 + s^{\text{(t)}}\right) \mu_k = 0,
\end{eqnarray}
where $k$ is the wavenumber, equal to the Euclidean norm of the wave vector $\vec{k}$ of the mode. 

It is important to note two facts in order to introduce a correct approximation to the background-dependent mass $s^{\text{(t)}}$. First, this mass is almost indistinguishable from the corresponding relativistic mass away from the region surrounding the bounce, the only region where the quantum geometry effects are relevant. Second, in a rough description of the background dynamics, the potential only has a significant role during inflation, while being almost negligible in the rest of the evolution.\\

\subsection{Initial conditions - Choice of a vacuum}

Before analytically approximating the background-dependent mass and solving the dynamics of the tensor modes, we need to specify the initial conditions for those modes. From the viewpoint of quantum field theory, fixing these initial data is equivalent to selecting a vacuum state. As mentioned in the Introduction, we here adopt the NO-AHD proposal, which selects a vacuum state adapted to the background, guarantees a good ultraviolet behavior and suppresses spurious oscillations in the PPS \cite{NMT}. In principle, this vacuum state can be determined in the following way. Any solution to the mode equations, suitably normalized (with respect to the Klein-Gordon product), can be expressed in the form \cite{NMT}
\begin{eqnarray} \label{NO-AHD modes}
    \mu_{k} = \sqrt{-\frac{1}{2 \textrm{Im}(h_{k})}}e^{i \int_{0}^{\eta}d\tilde{\eta} \textrm{Im}(h_{k})(\tilde{\eta})},
\end{eqnarray}
where $\eta$ is the conformal time and $h_{k}$ is a complex function that satisfies the Riccati equation
$\dot{h}_{k} = k^{2} + s^{\text{(t)}} + h_{k}^{2}$. Here $s^{\text{(t)}}$ denotes again the effective mass of the tensor perturbations. The NO-AHD vacuum is characterized by the function $h_k$ admitting the following asymptotic expansion in the ultraviolet limit:
\begin{equation}\label{asymptotich} 
    \frac{1}{h_k}\sim  \frac{i}{k}\left[1-\frac{1}{2k^2}\sum_{n=0}^{\infty}\left(\frac{-i}{2k}\right)^{n}\gamma_n \right],
\end{equation}
where the (background-dependent) coefficients $\gamma_{n}$ are independent of the wavenumber and determined by a(n asymptotic) recurrence relation derived from the condition that the Hamiltonian corresponding to the mode dynamics \eqref{MS_eq} becomes diagonal \cite{NMT}. In this relation, the first coefficient is given by the effective mass, $\gamma_{0} = s^{(t)}$.

It is possible to show that the resulting $h_{k}$ must have a negative imaginary part (at least asymptotically), so that Eq. \eqref{NO-AHD modes} provides positive-frequency modes \cite{NMT}. Moreover, the asymptotic expansion determines a unique solution for all $k$ in regimes where the inflaton potential is negligible or sufficiently small \cite{NM}. This solution leads to the desired initial conditions for our analytic integration.

\section{Analytic solutions}

The formulas presented in the previous section provide the evolution of the background and the effective mass in the mode equations. However, these equations do not admit analytic solutions over the entire background evolution. To overcome this problem and find analytic expressions, we divide the evolution of the background into three distinct periods: the bounce period, the kinetically dominated period, and the slow-roll inflationary period, considering instantaneous transitions between them (for further information, see Refs. \cite{NM, NMY}). 

\subsection{Bounce period}

The region around the bounce is characterized by the effective dynamics of LQC, which incorporates quantum corrections into the background. During this period of the evolution, and restricting the discussion, as we said, to situations of phenomenological interest, the value of the inflaton potential is negligible compared to the kinetic contribution to the energy density, resulting in a kinetic bounce. An exact analytic solution of the mode equation is not available in this region owing to the quantum corrections present in the background-dependent mass. Nevertheless, it has been proved that this effective mass can be well approximated by a Pösch-Teller (PT) potential \cite{Staro_LQC_1, Wu_2018}. Requiring that this potential coincides with the effective mass both at the bounce and at the end of the bounce period, when the quantum corrections to the relativistic expression of the mass are not important anymore, we obtain
\begin{equation}
    s_{\text{PT}}=\frac{U_0}{\cosh^2[\alpha (\eta-\eta_B)]},
\end{equation}
where $U_0=8\pi\rho_c/3$ and $\alpha=\text{arccosh}(a_0^2)/(\eta_0 - \eta_B)$. Here, $\eta_B$ is the conformal time at the bounce and $a_0$ denotes the scale factor at the end of the bounce period, that occurs at the conformal time $\eta_0$. 

To ensure that the relative error in approximating the effective mass using the PT approximation remains always below $25\%$, we fix the end of the bounce period at a proper time $t_0=0.4$, as discussed in Ref. \cite{NM}. For simplicity, we set  $\eta_0 =0$. The value of $\eta_B$ is then determined using the relation between proper and conformal times during the quantum epoch after the bounce (evaluated at the end of this period), 
\begin{equation} \label{conf-prop} 
   { \eta - \eta_\text{B} = {}_2 F_1\left(\frac{1}{6}, \frac{1}{2}, \frac{3}{2}; -24\pi\rho_c (t- t_\text{B})^2\right) (t-t_\text{B} )}.
\end{equation}
Here, ${}_2 F_1$ is the hypergeometric function. This leads to $\eta_B \approx -0.35$. The general solution of the tensor modes $\mu_k$ for the PT effective mass can indeed be obtained analytically. It has the form \cite{NM}
\begin{align} \label{mupt}
\mu_k^{\text{PT}}= &M_k[x(1-x)]^{-ik/2\alpha}{}_2 F_1\left(b^{(+)}_k,b^{(-)}_k;b_k;x\right)\nonumber \\+&N_k x^{ik/2\alpha}(1-x)^{-ik/2\alpha}{}_2 F_1\left(b^{(+)}_k-b_k+1,b^{(-)}_k-b_k+1;2-b_k;x\right),
\end{align}
where $x = [1- e^{-2\alpha(\eta - \eta_B)}]^{-1}$ and $M_k$ and $N_k$ are integration constants, which can be fixed giving initial conditions on the modes. In addition, the parameters of the hypergeometric function are
\begin{eqnarray}
    b_k=1-\frac{ik}{\alpha}, \quad b_k^{(\pm)}= b_k - \frac{1}{2} \pm  \sqrt{\frac{1}{4}+\frac{8\pi\rho_c}{3\alpha^2}}.
\end{eqnarray}

If we focus our attention exclusively on the bounce period, the NO-AHD vacuum state defined in the previous section corresponds in fact to the choice $M_k=\frac{1}{2k}$ and $N_k=0$ \cite{NM}, that we adopt in the following. We emphasize that this state is not adapted to the whole pre-inflationary and inflationary stages of the background evolution, but only to the bounce period. In fact, using the asymptotic expansion that characterizes this vacuum in the union of the three periods in which we divide the background evolution is not a viable procedure because we loose the desired analyticity in our calculations, among other reasons because the transitions between the different periods are instantaneous (and therefore not smooth). Restricting our considerations to the bounce period, we obtain a state that captures the relevant physics of this epoch, especially relevant for our study. Remarkably, starting from it, we will see later that it is possible to reach the genuine NO-AHD state by a Bogoliubov transformation \cite{NM}, performed once the perturbation modes have frozen after exiting the cosmological horizon during slow roll.

\subsection{Kinetic dominance}

The bounce period is followed by a classically relativistic period of kinetic dominance in the inflaton energy density. From this moment on, the residual quantum geometry effects can be ignored. Besides, the contribution of the inflaton potential is still negligible. Hence, it can be shown that the effective mass becomes \cite{NM}
\begin{eqnarray} \label{mass_KD} 
    s_\text{GR}^{(t)} (\eta) = \frac{1}{4} \left[\eta + \frac{1}{2a_0 H_0} \right]^{-2} ,
\end{eqnarray}
where $H_0 $ is the Hubble parameter at the end of the bounce period (and beginning of the kinetically dominated one), which we recall that has been taken at vanishing conformal time. This approximation is valid up to a certain conformal time, $\eta_\text{i}$, where the slow-roll inflationary period starts. We determine this conformal time $\eta_i$ by requiring that the relative error between the effective mass from LQC (\ref{mass_LQC}) and the mass (\ref{mass_KD}) (corresponding to a classical kinetic regime) remains below $25\%$. The general expression of the mode solutions during kinetic dominance can then be analytically computed, 
\begin{eqnarray}\label{mukin}
\mu_{k}^{\text{KD}} (\bar{y}) = \sqrt{\frac{\pi \bar{y}}{4}} \Big[C_k H_0^{(1)} (k\bar{y}) + D_k H_0^{(2)}(k\bar{y}) \Big ],
\end{eqnarray}
where $\bar{y}=\eta + 1/(2a_0 H_0) $, and $H_{0}^{(1)}$ and $H_{0}^{(2)}$ are the Hankel functions of the zeroth order and the first and second kinds, respectively. The integration constants, $C_k$ and $D_k$, are fixed by the solution $\mu_{k}^{{\text{PT}}}$ during the bounce period by requiring continuity of the modes up to their first time derivative at the matching time with the classical regime (that is, at vanishing conformal time). These constants must satisfy the (Klein-Gordon) normalization condition $|D_k|^2-|C_k|^2=1$ \cite{NM}. We can thus obtain the following expressions for them:
\begin{eqnarray} \label{C_D}
C_{k} &=&  \frac{1}{H_{0}^{(1)}(\hat{k})} \left[2\sqrt{\frac{k_0}{\pi}}\mu_{k}^{{\text{PT}}}(0) 
- D_{k} H_{0}^{(2)}(\hat{k})\right],\\
D_{k} &=& \frac{i}{2}\sqrt{\frac{\pi}{k_0}}\left[ k H_{1}^{(1)}(\hat{k}) \mu^{{\text{PT}}}_{k}(0) 
- \frac{k_0}{2}H_{0}^{(1)}(\hat{k})  \mu^{{\text{PT}}}_{k}(0)
+ H_{0}^{(1)}(\hat{k}) \mu^{\prime {\text{PT}}}_{k}(0) \right],
\end{eqnarray}
where $\hat{k}=k/k_0$ and $k_0= 2a_0H_0$.

\subsection{Slow-roll inflation}

The conformal time for the end of kinetic dominance and beginning of slow-roll inflation is, according to our previous discussion, given by $\eta_\text{i}$. From that moment on, the inflaton potential cannot be considered negligible, because it drives inflation. The slow-roll approximation is based on an expansion of the background equations in terms of the parameters
\begin{eqnarray}
 \varepsilon_\text{W} = \frac{1}{16\pi } \left(\frac{W,_\phi(\phi)}{W(\phi)}\right)^2, \quad \delta_\text{W} = \frac{1}{8\pi } \frac{W,_{\phi \phi} (\phi)}{W(\phi)}.
\end{eqnarray}
During slow roll, these parameters are approximately constant and much smaller than one. 

The mass during this period is given by the formulas \cite{Langlois, Baumann}
\begin{eqnarray}  \label{nu}
    s^{(t)}_{\text{SR}} = -\frac{\nu^2 - \frac{1}{4}}{(\eta_{e}-\eta)^2}, \hspace{2cm} \nu= \sqrt{\frac{9}{4}+ 3\varepsilon_\text{W}},
\end{eqnarray}
where $ \eta_\text{e}$ is the conformal time at the end of inflation. The associated mode solutions are
\begin{equation}
    \mu_{k}^{\text{SR}}(\eta) = \sqrt{\frac{\pi}{4}(\eta_{e}-\eta)} \left[A_k H_\nu^{(1)} \left[k(\eta_{e}-\eta)\right] + B_k H_\nu^{(2)} \left[k(\eta_{e}-\eta)\right]\right],
\end{equation} 
where the integration constants $A_k$ and $B_k$ are fixed by imposing continuity of the modes up to their first derivative at the matching time $\eta_i$. This leads to the relation
\begin{eqnarray}
A_{k} = &&-i\sqrt{\frac{\pi}{16}\Delta \eta } \left[ kH_{\nu+1}^{(2)}\left(k\Delta \eta \right) - kH_{\nu-1}^{(2)}\left(k\Delta \eta \right) - \frac{H_{\nu}^{(2)}\left(k\Delta \eta \right)}{\Delta \eta } \right] \mu_{k}^{\text{KD}}(\eta_{i}) +i\sqrt{\frac{\pi}{4}\Delta \eta }H_{\nu}^{(2)}\left(k\Delta \eta \right) \mu_{k}^{\prime \text{KD}}(\eta_{i}),
\end{eqnarray}
and a similar one for $B_{k}$ changing the global sign and the Hankel functions of second order by Hankel functions of first order. We have called $\Delta \eta = \eta_e -\eta_i$. Finally, normalization of the modes implies that $|A_k|^2-|B_k|^2=1$. 

\section{The primordial power spectrum for tensor perturbations}

We now proceed to the computation of the PPS for tensor perturbations, using both analytic and numerical methods. The approximations introduced in the previous section provide all the necessary ingredients to analytically calculate the PPS. For this, we evaluate the (observationally relevant) modes at $\eta_f$, the conformal time for which they become all frozen. In practice, this time can be defined as the instant where the mass term in Eq.\eqref{MS_eq} becomes much smaller than (the supremum of) $k^2$. 

The PPS is given by \cite{Baumann, Langlois}
\begin{equation} \label{PPS_def}
    \mathcal{P}_\mathcal{T}= \frac{32k^3}{\pi}\frac{|\mu_k^{\text{SR}}(\eta_{f})|^2}{a(\eta_{f})^2}.
\end{equation} \\

On the other hand, the window of observable modes approximately corresponds to the interval $ 2 \times10^{-4} {\rm Mpc}^{-1} \leq k/a_{today} \leq 6 \times 10^{-1} {\rm Mpc}^{-1}$, which therefore depends on the background cosmology through the present value of the scale factor $a_{today}$ \cite{Planck_2018}. With the convention that $a_{B}=1$, which we have adopted, it is clear that $a_{today}=e^{n_T}$, where $n_T$ is the number of e-folds from the bounce until now. Using that an inverse megaparsec is approximately $5 \times 10^{-58}$ inverse Planck lengths, the observable window becomes $[1,3\times10^3]\times10^{-61} e^{n_T}$. Moreover, for the phenomenologically interesting case of short-lived inflation in LQC, in which the traces of quantum effects on the primordial perturbations are not expected to be wiped out owing to a long duration of the inflationary period, numerical background studies carried out in previous works, for instance in Refs. \cite{AG1,Morris}, lead to a total number of e-folds in the range $130 \leq n_T \leq 143$. This order of magnitude is compatible with our consideration of around 70 e-folds until the end of inflation, along with the most recent proposals on the number of remaining e-folds from the end of inflation to the present, based on observations of large scale structures \cite{Zharov}. Motivated by these arguments and for the sake of concreteness, we will pay an especial attention to wavenumbers $k \in \left[10^{-4}, 10^2\right]$, which are potentially observable in scenarios with short-lived inflation according to our previous comments \cite{NM}.

Regardless of whether inflation is short-lived or not, for wavenumbers of this order (or smaller) in terms of the Planck scale, the argument in the Hankel functions of our mode solutions during slow roll turns out to be much smaller than one. Therefore, using the properties of the Hankel functions \cite{Abra}, we can rewrite the solution as 
\begin{eqnarray}
        |\mu_k^{\text{SR}}|^2 \simeq \frac{1}{4\pi} (\eta_{\text{e}}-\eta_{f}) |\Gamma(\nu_f)|^2 \left[\frac{k(\eta_{\text{e}}-\eta_{f})}{2}\right]^{-2\nu_f} |A_k - B_k|^2,
\end{eqnarray}
where $\Gamma$ is the gamma function and $\nu_f$ is the slow-roll parameter \eqref{nu} evaluated at $\eta_f$. The corresponding expression of the PPS is 
\begin{eqnarray} \label{eq_C_v_T}
        \mathcal{P}_\mathcal{T} (k) \simeq C_{\nu} k^{3-2\nu_f} |A_k-B_k|^2, \quad \text{where} \quad C_\nu = \frac{16 |\Gamma(\nu_f)|^2  }{\pi^2 a_{f}^2} \left(\frac{\eta_{e}-\eta_{f}}{2}\right)^{1-2\nu_f}.
\end{eqnarray}
Here, $a_f$ is the scale factor at the time where the modes have frozen. We note the dependence of the PPS on the integration constants of the modes, which are directly related to the choice of vacuum state. Hence, different choices lead to different spectra. This is especially so with respect to the oscillations induced by our approximations \cite{NM}, according to our comments at the end of Subsec. III.C. Indeed, in general, the spectrum may contain fast oscillations, even if the norms of the constants $A_k$ and $B_k$ vary slowly, owing to the influence of their respective phases $\theta_A$ and $\theta_B$. These oscillations pump power in average to the PPS. However, as we anticipated, it is actually possible to remove them by readjusting our choice of vacuum state with a convenient Bogoliubov transformation \cite{NM}, namely
 \begin{eqnarray}
A_k \to \Tilde{A}_k = |A_k|, \quad \quad B_k \to \Tilde{B}_k = |B_k|.
\end{eqnarray}
The final formula of the NO-AHD PPS is thus
\begin{eqnarray}
\mathcal{P}_\mathcal{T}(k) = C_\nu k^{3-2\nu} \left(|B_k| - |A_k|\right)^2 .\label{eq_PW_formula_cstes}
\end{eqnarray}
This expression retains the essential scale-dependent features of the original spectrum, excluding the superimposed rapid oscillations. Notably, it corresponds to the envelope of the minima of the oscillatory power spectrum.

In addition, for comparison, we want to compute the PPS by numerical integration of the evolution of the tensor modes. For this purpose, we use the \texttt{PYOSCODE} integrator \cite{Pyoscode}. This integrator is based on a \texttt{Runge-Kutta}-like stepping procedure that incorporates the Wentzel-Kramers-Brillouin (WKB) approximation to efficiently skip regions where the frequency of the solution varies slowly. This method allows for the computation of mode solutions with both high accuracy and significantly reduced computational time. Our numerical integration starts after the bounce period, just when the kinetically dominated period begins. Recall that we can determine explicitly the state of the perturbations selected by our vacuum prescription during the quantum epoch. Evaluating this state at the end of the bounce period, we obtain the initial data needed for the integration of the tensor modes. The numerical integration is performed over the whole evolution, until the modes become frozen during the inflationary era. The final step is to use the definition \eqref{PPS_def} of the PPS to get the spectrum obtained by our numerical analysis. For convenience, we normalize the PPS by absorbing the overall constant $C_\nu$ in Eq. \eqref{eq_C_v_T}.

\begin{figure}[h]
    \centering
    \includegraphics[width=0.8\textwidth]{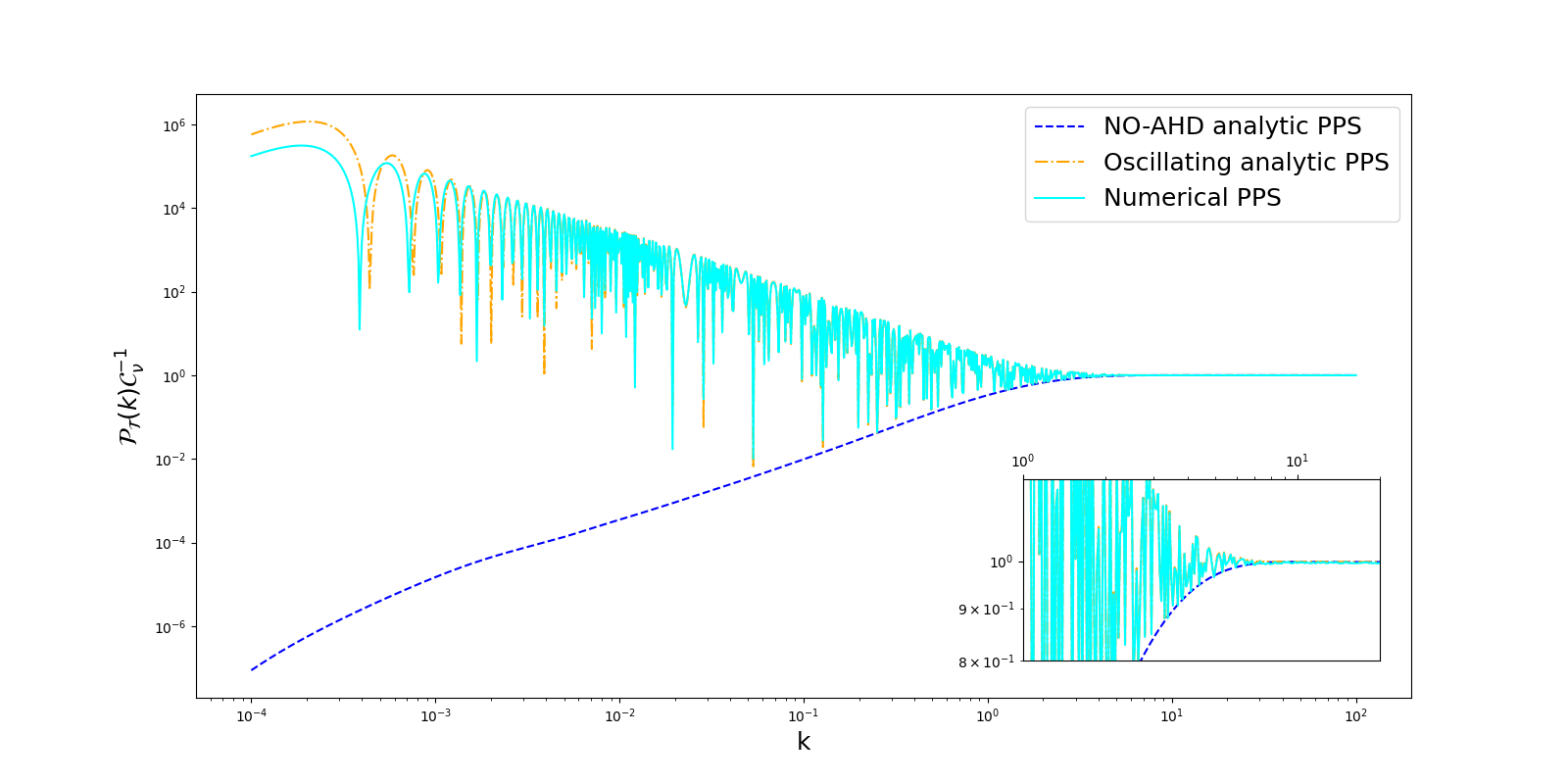} 
    \caption{Normalized PPS for short-lived inflation. We show the numerical computation of the oscillating PPS (continuous cyan line), the analytic approximation to this PPS (dot-dashed orange line), and the analytic approximation to the PPS of the NO-AHD vacuum (dashed blue line), obtained with a Bogoliubov transformation. We include an inset enlarging the region $1\leq k \leq 20$ (framed in the PPS) to see the details when power suppression appears. We have taken $\gamma=0.2375$ for the Immirzi parameter, $\phi_\text{B}=-1.45$ for the value of the inflaton at the bounce, positive inflaton time derivative at the bounce $\dot{\phi}_\text{B} >0 $, and considered a Starobinsky inflaton potential with mass $m=2.51 \times 10^{-6}$.}
    \label{fig:PPS_case1}
\end{figure}

First, we examine the analytic and the numerical PPS obtained using the NO-AHD criterion for short-lived inflation, with initial conditions $\phi_B = -1.45$ and $\dot{\phi}_B > 0$, as shown in Fig.\ref{fig:PPS_case1}. We observe that the analytic NO-AHD PPS derived from the final Bogoliubov transformation corresponds indeed to the envelope of the minima of the oscillating numerical PPS. The excellent agreement between our numerical results and the analytic approximation confirms that the latter captures all the essential physics. We only note a slight mismatch in the oscillation period in the infrared region, before applying the Bogoliubov transformation. This mismatch has no relevant consequences for the final NO-AHD PPS. Actually, a similar phenomenon was detected in Refs. \cite{MVY, MVY2} for the case of a quadratic inflaton potential. This discrepancy seems to originate in the transition epoch between the kinetically dominated regime and the beginning of inflation, where our analytic approximation becomes less accurate.

In the case of an inflaton with a mass term, it has been shown that correcting the approximation to the effective mass with contibutions of the inflaton potential is possible to improve the accuracy during the transition between kinetic dominance and slow roll \cite{NMY}. This method can also be implemented in the case of the Starobinsky potential, leading to improved results during the transition epoch (see Ref. \cite{SantoTomas} for a preliminary work on this topic), although the corresponding PPS does not differ appreciably from our findings once the spurious oscillations have been eliminated, as we have commented. 

\begin{figure}[h]
    \centering
    \includegraphics[width=0.8\textwidth]{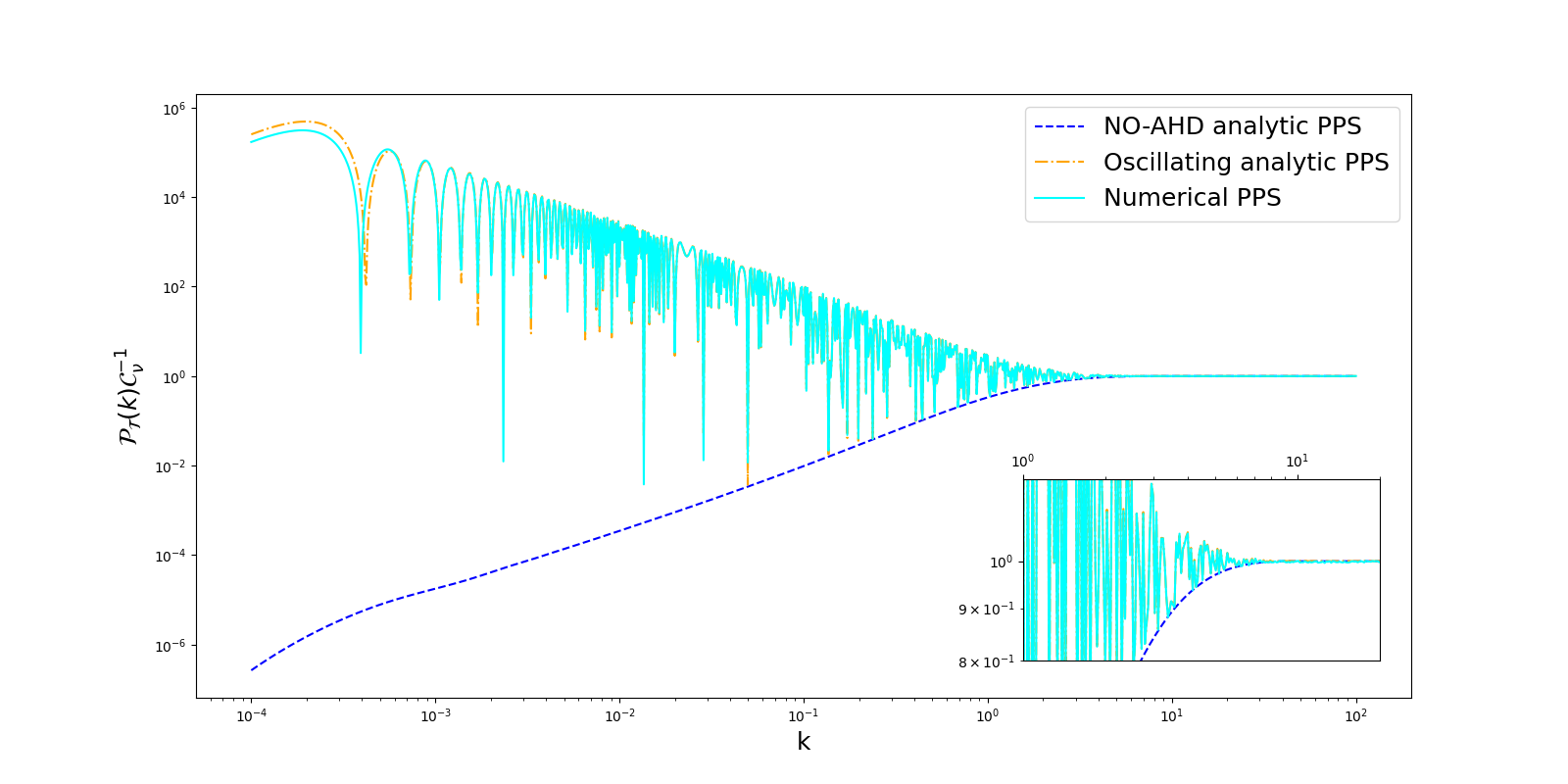} 
    \caption{Normalized PPS for long-lived inflation. We show the numerical computation of the oscillating PPS (continuous cyan line), the analytic approximation to this PPS  (dot-dashed orange line), and the analytic approximation to the PPS of the NO-AHD vacuum (dashed blue line), obtained with a Bogoliubov transformation. We include an inset enlarging the region $1\leq k \leq 20$ (framed in the PPS) to see the details when power suppression appears. We have taken $\gamma=0.2375$ for the Immirzi parameter, $\phi_\text{B}=0.97$ for the value of the inflaton at the bounce, positive inflaton time derivative at the bounce $\dot{\phi}_\text{B} >0 $, and considered again a Starobinsky inflaton potential with mass $m=2.51 \times 10^{-6}$.}
    \label{fig:PPS_case2}
\end{figure}

We have also studied another case of short-lived inflation, characterized by the initial conditions $\phi_B = 3.63$ and $\dot{\phi}_B < 0$. We obtain identical analytic and numerical PPS as before, with no significant deviation from the case presented in Fig. \ref{fig:PPS_case1}. 

\begin{figure}[h]
    \centering
    \includegraphics[width=1\textwidth]{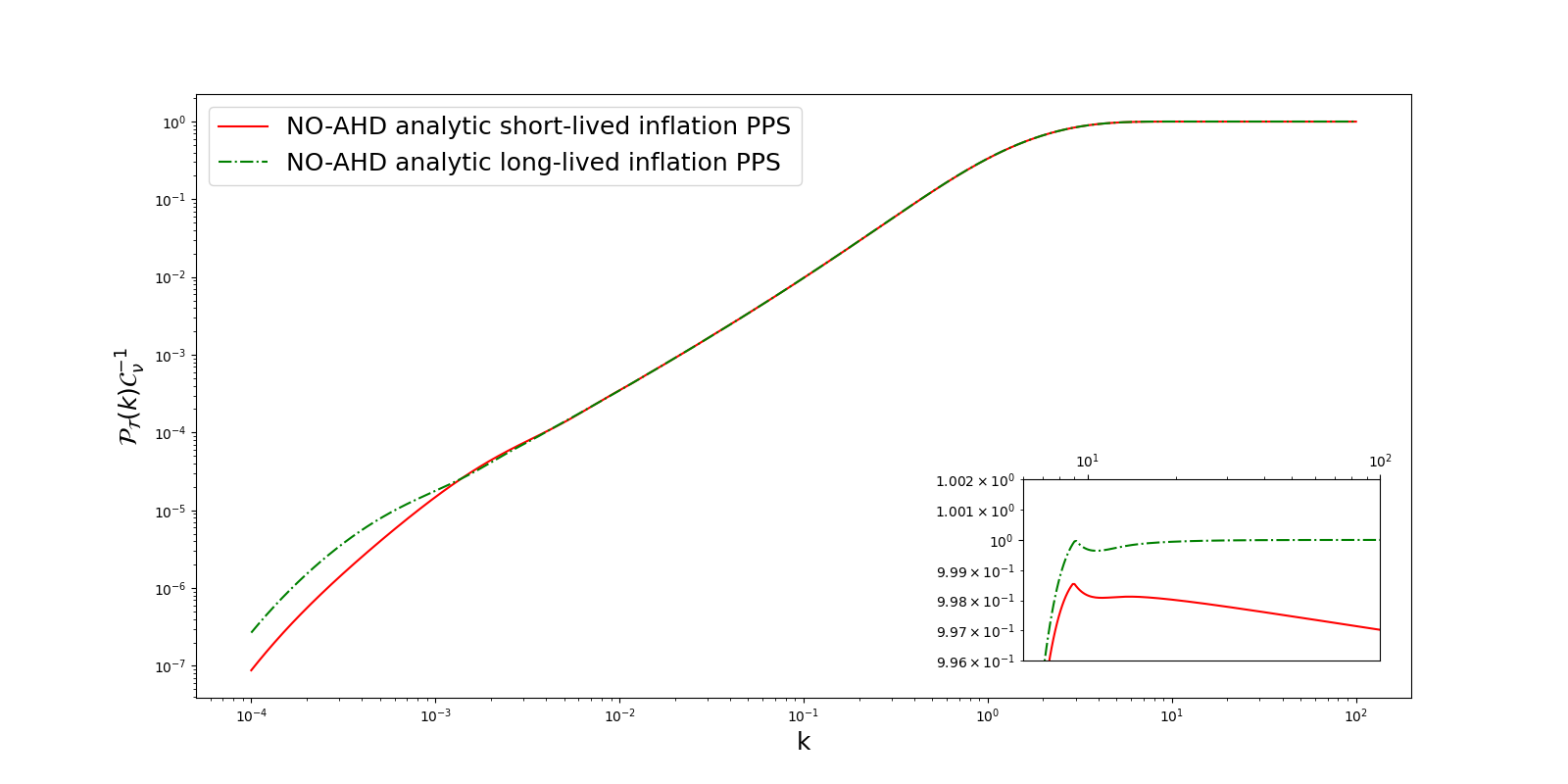} 
    \caption{Normalized analytic PPS for short- and long-lived inflation. The three PPS correspond to the analytic approximation to the PPS of the NO-AHD vacuum, including the final Bogoliubov transformation, for three different initial conditions on the inflaton and its derivative at the bounce. We show the analytic PPS for short-lived inflation with $\phi_\text{B}=-1.45$ and  $\dot{\phi}_\text{B} >0$ (red solid line), and the analytic PPS for long-lived inflation with $\phi_\text{B}=0.97$ and $\dot{\phi}_\text{B} >0$ (green dot-dashed line). We include an inset enlarging the region $6\leq k \leq 20$ (framed in the PPS) to see the different tilts in detail. We have taken $\gamma=0.2375$ for the Immirzi parameter and considered a Starobinsky inflaton potential with mass $m=2.51 \times 10^{-6}$.}
    \label{fig:all_PPS}
\end{figure}

Finally, we have analyzed a case with long-lived inflation. We have chosen initial conditions $\phi_B = 0.97$ and $\dot{\phi}_B > 0$, which lead to more than one million e-folds. The corresponding analytic and numerical PPS closely resemble those found in the short-lived inflationary scenario (see Fig. \ref{fig:PPS_case2}). Notably, the cutoff scale is indistinguishable in both scenarios, suggesting that it is directly determined only by the preinflationary background dynamics and the choice of vacuum state. This is further supported by the fact that the PPS is essentially identical in the infrared region except very far from the cutoff scale, where a slight discrepancy appears between the cases with short- and long-lived inflation. Actually, according to our computations, this difference can be reduced to negligible levels if we readjust the duration of the kinetic period by imposing a much smaller relative error between the masses \eqref{mass_LQC} and \eqref{mass_KD}. On the other hand, a relevant distinction arises at large scales, where the tilt, characterized by the slow-roll parameter $\nu$, differs between short- and long-lived inflation, as expected from the different values of the inflaton potential and its derivatives. In Fig. \ref{fig:all_PPS} we see that the value of $\nu$ in the long-lived inflationary case is closer to the de Sitter value than in the cases of short-lived inflation, as it is expected from a straightforward calculation using the definition of $\nu$.

\section{Conclusion}

In this work, we have computed the tensor PPS of a NO-AHD vacuum state in the hybrid LQC with a Starobinsky potential, considering both short- and long-lived inflationary scenarios. We have carried out this computation approximately by analytic means and also numerically, using different initial conditions at the bounce.

In both types of scenarios, we have calculated the analytic PPS by dividing the preinflationary and inflationary evolution in different periods and employing suitable approximations to the effective mass in those periods. With these approximations, it is possible to obtain the more general solution of the tensor mode equations in analytic form. During the preinflationary period, the Starobinsky potential is ignorable in practice for the considered background solutions, leading to a kinetic preinflationary regime. In a first part of this regime, following the big bounce that occurs in LQC, the quantum effects of the geometry are important, making the exact tensor mode equations analytically unsolvable. To circumvent this problem and find analytic solutions, we have employed a Pösch–Teller potential to approximate the effective mass. Once quantum effects become negligible in the background dynamics, the effective mass matches that of a classical kinetically dominated evolution in general relativity, for which the tensor mode equations admit also analytic solutions. Finally, after the two aforementioned periods, the potential becomes dominant and the Universe enters a slow-roll inflationary phase. During slow roll, the general solution of the tensor modes adopts a well known analytic expression in terms of Hankel functions.  

The preinflationary non-stationary epoch changes the background dynamics with respect to slow roll and invalidates the use of a Bunch-Davies state as a privileged vacuum, at least for modes with scales comparable to those that are characteristic of the preinflationary regimes. In the LQC context, several proposals have been developed to address the problem of selecting a preferred state to serve as the vacuum under these conditions. In this work, we have adopted the NO-AHD proposal, which is based on an asymptotic diagonalization of the Hamiltonian of the perturbations at large wavenumbers, which provides a natural identification of positive frequencies in that sector. In addition, the proposal selects a state which is optimally adapted to the background, resulting in slowly varying mode amplitudes that provide a non-oscillating PPS. To determine this NO-AHD state, we have first focused our attention on the quantum period after the bounce, capturing the physical phenomena characteristic of this epoch by selecting a state that diagonalizes the perturbative Hamiltonian in the ultraviolet. Such a state can be found analytically after using the aforementioned Pösch–Teller approximation for the effective mass. The evaluation of the modes of this state and their derivatives at the beginning of the classical kinetically dominated period have provided initial condition for the integration of the tensor perturbations once the quantum effects have diluted. Spurious oscillations in the PPS arising from our approximations have been removed by a suitable Bogoliubov transformation after horizon crossing, when all relevant modes have frozen. This transformation gives the genuine NO-AHD PPS. While the NO-AHD vacuum state had previously been studied for quadratic potentials both in the hybrid and dressed metric approaches to LQC \cite{NM,MVY,MVY2,AMV}, its implementation in the Starobinsky scenario within LQC is entirely novel.

In addition to the approximated, analytic methods used to compute the PPS, we have also numerically integrated the evolution of the tensor modes for the same set of background initial conditions, obtaining the corresponding PPS and comparing the results. The main conclusion is that our results confirm the extraordinary accuracy of the analytic approximations, and that they are valid in a remarkably wide range of situations which include both short- and long-lived inflationary scenarios, with a change in the number of e-folds of more than five orders of magnitude. In both cases, the resulting analytic and numerical PPS are nearly identical, with the same cutoff scale. The emergence of this scale, independent of the details of the inflationary period, seems to be directly connected to the preinflationary dynamics of the background in LQC near the bounce, combined with the choice of a vacuum state that is able to encode the quantum physics of these dynamics. The only remarkable difference between the PPS in the different scenarios that we have considered lies in their spectral tilt for large wavenumbers, a behavior which is fully explained by the value and form of the inflaton potential during slow roll (via the slow-roll parameters). Consistent with these comments, we find a tilt that is closer to the de Sitter value for long-lived inflation.

Together with previous results about the quadratic potential, this work provides a strong support for the use of our analytic approximations to describe the PPS of a NO-AHD vacuum in LQC, independently of the specific form of the inflaton potential and for an ample variety of scenarios, as far as the bounce remains kinetically dominated for them. These scenarios allow for both short- and long-lived inflation. In this way, our results open new avenues for future parametrizations of the PPS, enabling Bayesian analyses of the CMB that would permit the confrontation of the predictions of LQC with observational data (either eventually accessible in the case of tensor perturbations, or even available for scalar perturbations, which can be treated in a similar manner).

\acknowledgments

The authors are thankful to B. Elizaga Navascu\'es, P. Santo-Tomás Ros, and J. Y\'ebana-Carrilero for helpful conversations. This work was partially supported by MCIU/AEI/10.13019/501100011033 and FSE+ under the Grant No. PID2023-149018NB-C41.

\end{document}